\documentclass{article}
\usepackage{arxiv}
\usepackage{natbib}
\usepackage{graphicx}
\usepackage{multirow}
\usepackage{subcaption}
\usepackage{booktabs}
\usepackage{makecell}
\usepackage{todonotes}
\usepackage{lmodern,url}
\usepackage{siunitx}
\usepackage{xspace}
\usepackage{soul}
\usepackage{gensymb}

\usepackage{lscape}

\title{Geometric deep learning assists protein engineering. Opportunities and Challenges}

\usepackage{authblk}

\author[1,2]{Julián García-Vinuesa}
\author[3]{Jorge Rojas}
\author[3]{Nicole Soto-García}
\author[1,2]{Nicolás Martínez}
\author[3,4]{Diego Alvarez-Saravia}
\author[2,3]{Roberto Uribe-Paredes}
\author[5]{Mehdi D. Davari}
\author[2,6]{Carlos Conca}
\author[1,2]{Juan A. Asenjo}
\author[3,5\thanks{\tt{david.medina@umag.cl}}]{David Medina-Ortiz}

\affil[1]{Departamento de Ingeniería Química, Biotecnología y Materiales, Universidad de Chile, Beauchef 851, Santiago, Chile}
\affil[2]{Centre for Biotechnology and Bioengineering, CeBiB, Universidad de Chile, Beauchef 851, Santiago, Chile}
\affil[3]{Departamento de Ingeniería en Computación, Universidad de Magallanes, Avenida Bulnes 01855, Punta Arenas, Chile}
\affil[4]{Centro Asistencial de Docencia e Investigación, CADI, Universidad de Magallanes. Av. Los Flamencos 01364, Punta Arenas, Chile}
\affil[5]{Leibniz-Institute of Plant Biochemistry Department of Bioorganic Chemistry,  Weinberg 3, D-06120 Halle, Germany}
\affil[6]{Center for Mathematical Modeling, (CMM) (UMI CNRS 2807), Department of Mathematical Engineering, Universidad de Chile, Beauchef 851, Santiago, Chile}

\date{}

\renewcommand{\headeright}{ }
\renewcommand{\undertitle}{ }

\begin{document}
\maketitle

\begin{abstract}
Protein engineering is experiencing a paradigmatic shift through the integration of geometric deep learning into computational design workflows. While traditional strategies—such as rational design and directed evolution—have enabled relevant advances, they remain limited by the complexity of sequence space and the cost of experimental validation. Geometric deep learning addresses these limitations by operating on non-Euclidean domains, capturing spatial, topological, and physicochemical features essential to protein function.

This perspective outlines the current applications of GDL across stability prediction, functional annotation, molecular interaction modeling, and \textit{de novo} protein design. We highlight recent methodological advances in model generalization, interpretability, and robustness, particularly under data-scarce conditions. A unified framework is proposed that integrates GDL with explainable AI and structure-based validation to support transparent, autonomous design. As GDL converges with generative modeling and high-throughput experimentation, it is emerging as a central technology in next-generation protein engineering and synthetic biology.
\end{abstract}


\section{Introduction}

Protein engineering is one of the most relevant strategies of modern biotechnology, enabling the design and optimization of proteins with desirable functions and properties for applications across medicine, industry, and synthetic biology \citep{poluri2017biotechnological}. Traditional strategies, such as rational design and directed evolution have led to significant advances \citep{liu2019state}. Nevertheless, these methods remain constrained by the immense size of protein space and the laborious demands of experimental validation \citep{vidal2023primer}.

Advances in machine learning (ML) have begun to reshape the landscape of protein engineering, offering data-driven strategies to predict relevant properties, including protein stability \citep{musil2024fireprot}, binding affinities \citep{siebenmorgen2020computational}, and catalytic properties \citep{chen2022recent}. Moreover, strategies like ML-guided directed evolution integrates predictive models with experimental screening, reducing the number of required iterations to discover optimal variants \citep{yang2019machine}.

The emergence of pretrained protein large language models has further enhanced ML strategies to aid protein design by capturing evolutionary and structural patterns that inform sequence annotation and optimization \citep{sarumi2024large}. In parallel, generative models—including variational autoencoders \citep{hawkins2021generating}, generative adversarial networks \citep{lin2022novo}, and diffusion-based architectures \citep{li2025thermodynamics}—have facilitated the \textit{de novo} design of proteins, while structure-aware methods leverage three-dimensional information to refine predictions.

Among structure-based approaches, geometric deep learning (GDL) has emerged as a promising framework. By encoding spatial and topological relationships through graph neural networks (GNNs) and equivariant architectures, GDL can captures the complex geometry of protein structures with high fidelity \citep{atz2021geometric}. These structural properties address key limitations of traditional ML models, which often reduce proteins to oversimplified representations that overlook allosteric regulation \citep{sheik2020integrated}, conformational flexibility \citep{teague2003implications}, and solvent-mediated interactions \citep{avery2022protein}.

Despite the progress in GDL, several critical challenges persist. Accurately capturing dynamic structural rearrangements \citep{Atz2021}, managing the scarcity of high-quality annotated datasets \citep{barbero2024addressing}, and enhancing model interpretability continue to limit broader adoption \citep{medina2024interpretable}. Many deep learning architectures still function as \textit{black boxes}, impeding the extraction of mechanistic insights and constraining their utility in guiding experimental design \citep{shwartz2017opening}. Even GDL models, despite their greater representational capacity, face significant limitations in capturing the full spectrum of protein conformational dynamics and environmental influences \citep{audagnotto2022machine}.

In this perspective, we present the recent advances in GDL for protein design, with a focus on how interpretability and structural awareness can be synergistically combined to enable transparent, efficient, and biologically informed design workflows. Rather than introducing a new model architecture, we aim to consolidate existing strategies into a cohesive pipeline—from structure acquisition and graph construction to training, interpretation, and experimental validation.

First, we review foundational concepts in protein representation and graph-based modeling, followed by a critical overview of state-of-the-art of GDL-based ML models and their applicability for predictive and generative protein design tasks. We then highlight the emerging role of explainable artificial intelligence (XAI) in guiding design decisions and improving biological interpretability. Finally, we propose a unified pipeline that integrates these components, bridging computational advances with experimental feasibility. Our goal is to provide a roadmap for researchers aiming to build interpretable, structure-aware models for protein engineering, and to highlight open challenges that must be addressed to realize their full potential.

\section{Geometric Deep Learning: Foundations, Key Properties, and Emerging Challenges}

Recent advances in computational infrastructure and the explosion of large-scale biological datasets have catalyzed the development of ML-based methods tailored to non-Euclidean domains \citep{Cao2020}. However, the increasing dimensionality of biological data poses critical challenges, commonly referred to as the \textit{curse of dimensionality}. As the number of dimensions grows, the volume of the feature space increase exponentially, resulting in sparse data distributions that compromise learning efficiency and impair model generalization \citep{Petrini2023}.

GDL has emerged as a foundational framework to address these challenges, enabling the modeling of structured data defined over graphs and three-dimensional molecular surfaces \citep{isert2023structure}. By aligning model architectures with the geometry of the data, GDL mitigates high-dimensional complexity through foundational principles, including symmetry and scale separation \citep{Bronstein2021}.

Symmetry, in this context, refers to a model's equivariance or invariance under specific group transformations. For protein structures, these include spatial manipulations such as rotations, translations, reflections, and permutations—operations that must preserve the physical validity of molecular geometry \citep{satorras2021n}. Architectures that respect these symmetries—particularly those equivariant to the Euclidean group $E(3)$ or the special Euclidean group $SE(3)$—have demonstrated fidelity in capturing the geometry and conformational behavior of biomolecules \citep{li2025thermodynamics}.

Complementing symmetry, scale separation allows complex biological signals to be decomposed into multi-resolution representations, often via wavelet-based filters or hierarchical pooling mechanisms \citep{Bronstein2021}. This facilitates the capture of both fine-grained residue-level interactions and long-range structural dependencies, both of which are critical for predicting molecular function and catalytic and structural properties \citep{wang2024graph}.

Together, these principles define the blueprint of modern GDL architectures—composed of equivariant linear layers, nonlinear activation functions, and invariant pooling operations. Such models have demonstrated state-of-the-art performance across a range of domains—from computer vision and quantum chemistry to structural biology—by enabling the extraction of robust, generalizable representations from geometrically structured inputs \citep{Ahmetoglu2022}.

Nevertheless, most current GDL frameworks rely on static or single-conformation protein representations, limiting their capability to capture functionally relevant conformational ensembles, allosteric transitions, or intrinsically disordered regions \citep{wang2024graph}. To address this, emerging strategies have begun to integrate dynamic information directly into geometric learning pipelines. 

Molecular dynamics (MD) simulations, for instance, provide atomistic trajectories that sample conformational diversity, enabling the construction of ensemble-based graphs or time-averaged features \citep{dror2012biomolecular}. Alternatively, multi-conformational graphs built from MD snapshots can capture flexible residue–residue contacts and fluctuating interaction networks. Some models further incorporate flexibility-aware priors—such as B-factors, backbone torsion variability, or disorder scores—into node and edge embeddings \citep{cao2024mcpnet}. These developments mark a critical shift toward probabilistic and dynamic GDL models that more faithfully reflect the fluid, adaptive nature of protein structures. Capturing cryptic binding pockets, transient interactions, and disordered segments will be essential for improving biological fidelity and enabling rational design \citep{hu2024exploring}.

A complementary challenge lies in the ability to generalize across protein families, functional contexts, and structural regimes. To this end, transfer learning strategies are increasingly employed to repurpose pretrained geometric models—often trained on large-scale structural datasets—for downstream tasks in protein engineering \citep{GVP-GNN}. Fine-tuning these models under data-scarce conditions has yielded relevant improvements in predictive performance and sample efficiency \citep{hu2019strategies}. Moreover, transferable protein embeddings—encoding spatial topology and biochemical context—can be repurposed across tasks; for instance, embeddings optimized for stability prediction may inform function classification or ligand-binding affinity estimation \citep{stark2022equibind}. These approaches support modular, interpretable GDL workflows, promoting the reuse and adaptation of computational priors across diverse biological problems. Systematic benchmarking of transferability across tasks, domains, and evolutionary clades remains an open challenge, with significant implications for the robustness, scalability, and translational impact of GDL-based approaches \citep{notin2022tranception}.

\section{Implementing Predictive Models Using Geometric Deep Learning Strategies} \label{sec:sec-implementing}

Developing predictive models for protein design requires the integration of structural, functional, and spatial information to faithfully capture biochemical behavior \citep{kyro2025model}. While early approaches often treated proteins as linear sequences of amino acids, recent advances in GDL have enabled models to directly incorporate three-dimensional molecular structures, significantly improving both biological relevance and predictive accuracy \citep{liu2024all}.

\begin{figure}[!htpb]
    \centering
    \includegraphics[width=\linewidth]{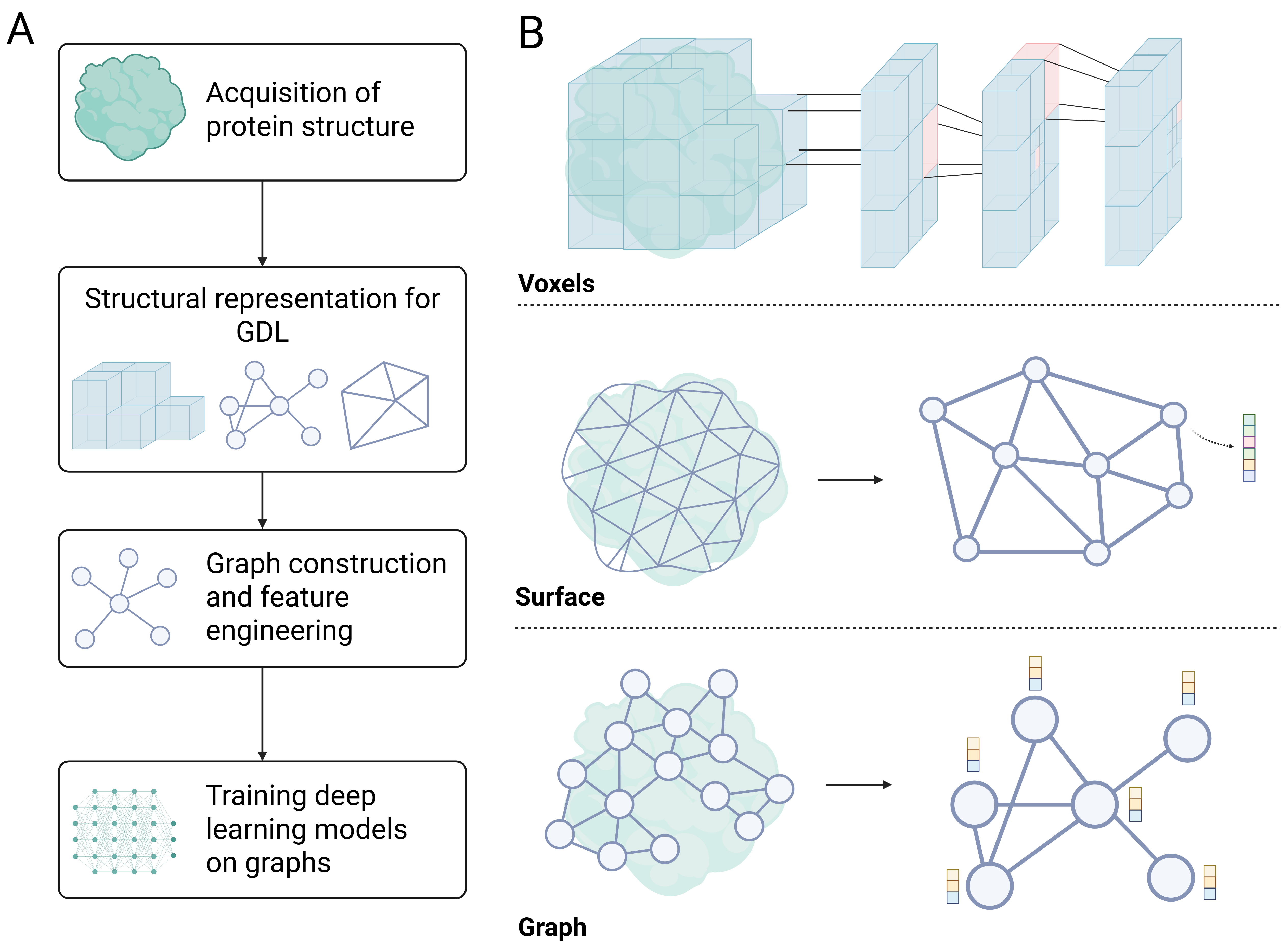}
    \caption{\textbf{Overview of the methodological pipeline and structural input strategies used to develop geometric deep learning models for protein prediction tasks.} \textbf{A} Schematic representation of the key steps involved in developing geometric deep learning (GDL)-based predictive models. The pipeline begins with the collection of protein structures from experimental databases or through structure prediction tools. These structures are then processed to generate suitable structural representations for GDL. Subsequently, graphs are constructed and enriched through feature engineering to produce model-ready inputs. Finally, predictive models are trained using selected graph neural network architectures, with performance evaluated using standard validation strategies and metrics. \textbf{B} Traditional strategies for representing protein structures as input for GDL-based models. These include voxel-based representations, which discretize 3D space into regular grids; surface-based representations, which model the molecular interface using meshes or point clouds; and graph-based representations, where atoms or residues serve as nodes and edges encode spatial or chemical interactions. }
    \label{fig:fig01}
\end{figure}

To illustrate how GDL strategies can be effectively applied, the following sections detail the core stages involved in constructing predictive models based on geometric representations. From the acquisition and preprocessing of protein structures to the development of deep learning architectures capable of interpreting complex topologies, each step plays a critical role in modeling molecular complexities and generating biologically meaningful predictions (Figure \ref{fig:fig01}~A).

\subsection{Acquisition of Protein Structures}

The foundation of any GDL-based model for protein analysis lies in the accurate acquisition of three-dimensional structures \citep{meng2025mvgnn}. Traditionally, structural biology has relied on high-resolution experimental techniques such as X-ray crystallography, nuclear magnetic resonance spectroscopy, and cryo-electron microscopy, which provide detailed spatial representations of protein conformations. However, these methods are often time-consuming, expensive, and limited in throughput \citep{bertoline2023before}. 

To address these limitations, advances in computational structural biology have led to the emergence of structure prediction tools like AlphaFold3 \citep{abramson2024accurate}, RoseTTAFold \citep{baek2021accurate}, and ESMFold \citep{lin2023evolutionary}. These tools, trained on vast protein datasets, offer accurate structural estimations from sequence data alone and have significantly democratized access to high-quality structural information, enabling their seamless integration into GDL pipelines \citep{wu2023geometric}.

\subsection{Structural Representations for GDL}

Once the protein structure is available, selecting an appropriate representation becomes critical, as it determines how spatial and physicochemical information is encoded and interpreted by the model \citep{musil2021physics}. Figure \ref{fig:fig01}~B summarizes the traditional strategies to represent protein structural information to develop GDL-based models. 

Volumetric grids, or 3D voxel-based representations, discretize the molecular structure into uniform spatial units, allowing each voxel to carry rich descriptors such as atomic density, hydrophobicity, and electrostatic potential \citep{Liu2024}. Despite their expressive power, voxel-based models are computationally expensive, and their memory demands can limit scalability \citep{cao2022geometric}. Techniques like sparse representations and resolution-adaptive grids have been proposed to mitigate these issues \citep{diaz2025learning}.

Alternatively, surfaces or molecular surfaces provide a more geometrically intuitive representation by modeling the protein as a triangulated mesh or a point cloud, capturing the contours and local topology of its exterior \citep{isert2023structure}. Each vertex can be annotated with geometric descriptors like curvature and shape index, as well as chemical properties such as hydropathy or charge distribution \citep{MaSIF}. This representation is particularly valuable for capturing surface-level interactions and has been extensively used in tasks involving ligand binding prediction or interface detection \citep{dai2021protein}.

Perhaps the most flexible and widely adopted representation in GDL is the three-dimensional graph \citep{isert2023structure}. In this strategy, proteins are modeled as graphs where nodes correspond to residues or atoms, and edges represent physical or spatial interactions, such as covalent bonds, hydrogen bonds, or proximity below a specified threshold \citep{Liu2024}. Graph-based representations allow the encoding of complex relationships between structural components without assuming a fixed grid, making them computationally efficient and naturally suited for neural architectures like GNNs \citep{Jumper2021IPA}. These graphs can incorporate diverse features on nodes—such as sequence embeddings, structural positions, and biochemical properties—and edge features capturing distance, directionality, and bond types, thus offering an integral molecular representation \citep{10.1093/bioinformatics/btac479}.

\subsection{Graph Construction and Feature Engineering}

Constructing a protein graph involves defining a comprehensive set of nodes and edges that reflect meaningful molecular relationships \citep{madani2024}. Residues or atoms are typically designated as nodes, enriched with features derived from both sequence and structural data \citep{Xia2021}. 

Traditional feature engineering strategies may include one-hot encodings, physicochemical descriptors, position-specific scoring matrices, and contextual embeddings generated by pretrained protein large language models like ESM \citep{HARDINGLARSEN2024108459}. The spatial arrangement is preserved through the inclusion of three-dimensional coordinates extracted from structural files, while  biological descriptors, such as solvent accessibility and secondary structure annotations, contribute to the functional characterization of each node \citep{MaSIF}.

Edges in the graph are defined by spatial or chemical proximity, often based on a distance cutoff (e.g., 6 Å) between atoms or residues \citep{jha2022prediction}. Advanced graph construction techniques incorporate additional interaction types, including disulfide bridges, hydrogen bonds, salt bridges, and $\pi-\pi$ stacking interactions \citep{xie2025ppi}. Tools like Graphein automate this process, assigning edge labels and weights based on interaction strength or biological relevance \citep{jamasb2022graphein}. This allows for the creation of highly informative and topologically consistent graphs, tailored to capture the subtleties of protein architecture and interaction networks.

\subsection{Training Deep Learning Models on Graphs}

With the graph representation defined, the next step in GDL-based protein modeling involves selecting a suitable learning architecture. Graph neural networks offer a robust framework for learning on non-Euclidean domains by propagating and aggregating information across local neighborhoods within the graph structure \citep{Jumper2021IPA}. At their core, GNNs implement a message-passing paradigm in which each node iteratively updates its feature vector by integrating information from its neighbors through a series of learnable transformations and aggregation functions. This iterative refinement allows the network to capture intricate, nonlinear dependencies encoded within a protein’s three-dimensional topology, making GNNs particularly well-suited for biomolecular modeling \citep{Shao2024}.

\subsubsection{Training strategies and selecting suitable GNN-based architectures}

\begin{figure}[!htpb]
    \centering
    \includegraphics[width=\linewidth]{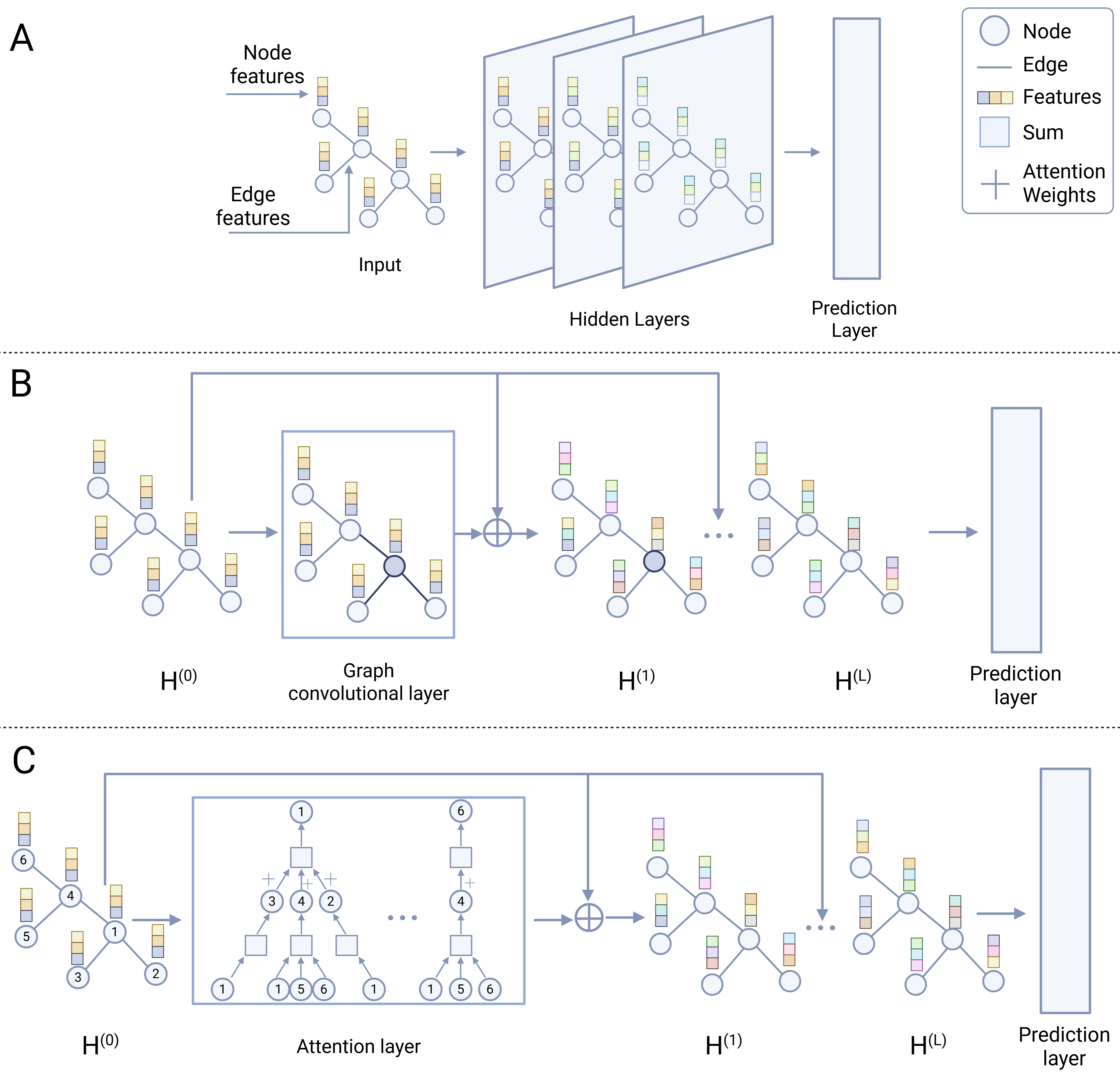}
    \caption{\textbf{Representative graph neural network (GNN) architectures for protein modeling tasks}.  Graph neural networks have emerged as powerful tools to learn on molecular and structural representations of proteins, capturing spatial, topological, and physicochemical features from their graph-based encodings. Shown are three widely used GNN variants. (Top) A classical message-passing neural network, which iteratively aggregates information from neighboring nodes to update node representations. (Middle) A graph convolutional network (GCN), which applies learnable convolutional filters over local node neighborhoods to integrate structural and biochemical context. (Bottom) A graph attention network (GAT), which enhances message passing by assigning attention weights to each neighboring node, allowing the model to differentially prioritize relevant interactions. Each architecture produces enriched node or graph-level embeddings used for downstream prediction tasks.}
    \label{fig:fig2}
\end{figure}

Figure~\ref{fig:fig2} describes representative GNN architectures widely used for protein-related prediction tasks. These include traditional GNN architectures based on message-passing strategies, graph convolution-based strategies (GCNs), and graph attention-driven mechanisms (GAT). Each architecture introduces distinct inductive biases and computational trade-offs, and their selection often depends on dataset scale, task complexity, and computational resources.

Training strategies must be carefully adapted to both the size of the dataset and the complexity of the underlying architecture. Full-graph training—where the entire molecular graph is processed as a whole—offers a global and coherent view of protein topology, but it is computationally expensive, particularly for large biomolecules or when scaled to batched workflows \citep{Bajaj2024}. To address these limitations, mini-batch training partitions the molecular graph into localized subgraphs, enabling memory-efficient and scalable learning in high-throughput settings \citep{Balaji2025}.

The choice of model—whether discriminative, generative, or explainable—should reflect the objectives of the task and the interpretability requirements. Discriminative models such as graph neural network classifiers or regressors are well suited for tasks with labeled datasets \citep{jia2025review}. 

In contrast, generative models are employed for structure generation, fitness landscape exploration, and \textit{de novo} protein design. Among these, Graph Variational Autoencoders (VGAEs) offer a probabilistic framework by learning latent representations from protein graphs through encoder–decoder architectures \citep{Kipf2016VGAE}. These models introduce stochasticity in the latent space, supporting uncertainty estimation and semi-supervised learning—features particularly valuable in low-data or exploratory design scenarios \citep{wang2022bayestab}. 

\subsubsection{Challenges on selecting GNN architectures}

Explainability is equally crucial in biomedical applications. Post hoc interpretation methods and intrinsically interpretable models are increasingly integrated for tasks such as mutational effect analysis, biomarker discovery, or rational regulatory design \citep{nandan2025graphxai}. Hybrid strategies, combining generative backbones with discriminative outputs and interpretability layers, are gaining traction for their ability to balance accuracy, transparency, and adaptability across tasks.

Architectural decisions embed inductive biases that can enhance or reduce task performance \citep{battaglia2018relational}. For instance, Graph Convolutional Networks assume smoothness and local homogeneity in feature space, a favorable property for tasks like domain classification or fold prediction \citep{min2020scattering}. Graph Attention Networks, by contrast, dynamically assign importance weights to neighboring nodes, allowing the capture of asymmetric or long-range dependencies that are critical for identifying non-contiguous active sites or allosteric interactions \citep{baranwal2022struct2graph}. Variational architectures such as VGAEs extend this representational flexibility by encoding protein graphs into probabilistic latent spaces, enabling uncertainty modeling and latent exploration—both of which are essential for protein design and structure-function inference \citep{mansoor2024protein}.

Generative graph models—particularly GAEs and VGAEs—extend beyond prediction to support data augmentation, structure completion, and rational protein design. By mapping proteins to latent spaces that encode biochemical and topological priors, these models allow interpolation between known structures and exploration of novel, designable variants \citep{Kipf2016VGAE}. Importantly, their probabilistic formulation supports uncertainty estimation, which is crucial for guiding wet-lab validation and early-stage drug discovery \citep{Genie}. Together, these generative and interpretable frameworks are transforming GDL into a flexible and modular foundation for next-generation protein engineering.

\subsubsection{Employed metrics for performance evaluation}

The performance of these models are typically evaluated using classification metrics—such as accuracy, precision, recall, F1-score, or ROC-AUC—for tasks involving discrete outputs, particularly under class imbalance \citep{medina2024protein}. Regression tasks, such as predicting $\Delta \Delta G$, thermostability, or solubility, rely on metrics including mean squared error (MSE), root mean squared error (RMSE), mean absolute error (MAE), and Pearson or Spearman correlation coefficients \citep{medina2020development}. 

Generative models within geometric deep learning frameworks are commonly evaluated through a combination of statistical, structural, and functional criteria. These may include log-likelihood, Kullback-Leibler (KL) divergence, and Frechet distance in latent space, as well as metrics assessing the validity, novelty, and diversity of generated structures \citep{shlens2014notes}. When applied to proteins or molecules, additional evaluations may involve 3D structural similarity (e.g., TM-score or RMSD), docking or binding affinity scores, and biophysical plausibility of generated candidates. Careful selection of evaluation metrics ensures that performance is both task-relevant and biologically meaningful \citep{Ingraham2019}.

\subsubsection{Generalization and Benchmarking}

Robust generalization in GDL workflows depends critically on dataset partitioning and validation. Random or stratified splitting may suffice in simple tasks, but for biological applications, homology-aware clustering is necessary to avoid information leakage due to evolutionary redundancy \citep{dwivedi2023benchmarking}. This is particularly important in tasks requiring cross-family generalization or fold novelty. Nested cross-validation frameworks, combined with hyperparameter tuning via grid search or Bayesian optimization, further safeguard against overfitting \citep{bischl2023hyperparameter}. Regularization strategies—such as dropout, weight decay, and early stopping—are widely applied to improve generalization, especially in low-N settings \citep{srivastava2014dropout}.

The emergence of benchmark datasets has been critical in standardizing evaluation across studies. For instance, TAPE provides a suite of sequence- and structure-informed tasks for benchmarking predictive models \citep{rao2019evaluating}. SABDab offers curated antibody structures to support interface and affinity predictions \citep{schneider2022sabdab}, while ProteinGym delivers mutational effect scans with experimentally validated $\Delta \Delta G$ data for stability assessments \citep{notin2023proteingym}. These datasets have enhanced comparability and reproducibility across GDL applications in structural biology.

\section{Geometric Deep Learning Applications for protein engineering tasks}

The design and optimization of proteins remain fundamental challenges in biotechnology and synthetic biology \citep{Khakzad2023}. These challenges are associated with the inherently high-dimensional nature of protein data, which integrates linear amino acid sequences with intricate three-dimensional conformations \citep{CARBonAra}. Capturing the relationships between sequence, structure, and function requires modeling approaches that can effectively navigate this complex, multi-scale landscape.

Traditional computational frameworks often apply linear sequence encodings or simplified two-dimensional graph-based representations, which may fail to capture essential spatial constraints and physicochemical interactions \citep{zhang2023}. In contrast, GDL enables the models to incorporate rich geometric and topological information directly into the learning process \citep{GVP-GNN}. By leveraging the three-dimensional nature of protein structures, GDL approaches can more accurately model the spatial dependencies related to biological functions and molecular interaction \citep{meng2025mvgnn}.

Recent developments have highlighted the utility of GDL in a wide range of protein modeling tasks, including: i) predicting interactions between proteins, ligands, and nucleic acids \citep{PeSTo, moon2024pignet2}; ii) modeling protein–protein binding interfaces \citep{zhu2025identifying}; iii) assessing physicochemical properties and inferring biological function \citep{Yuan2024GPSFun}; and iv) enabling \textit{de novo} protein design through structure-conditioned or sequence-generative models \citep{Rfdiffusion}.

\begin{figure}[!htpb]
    \centering
    \includegraphics[width=\linewidth]{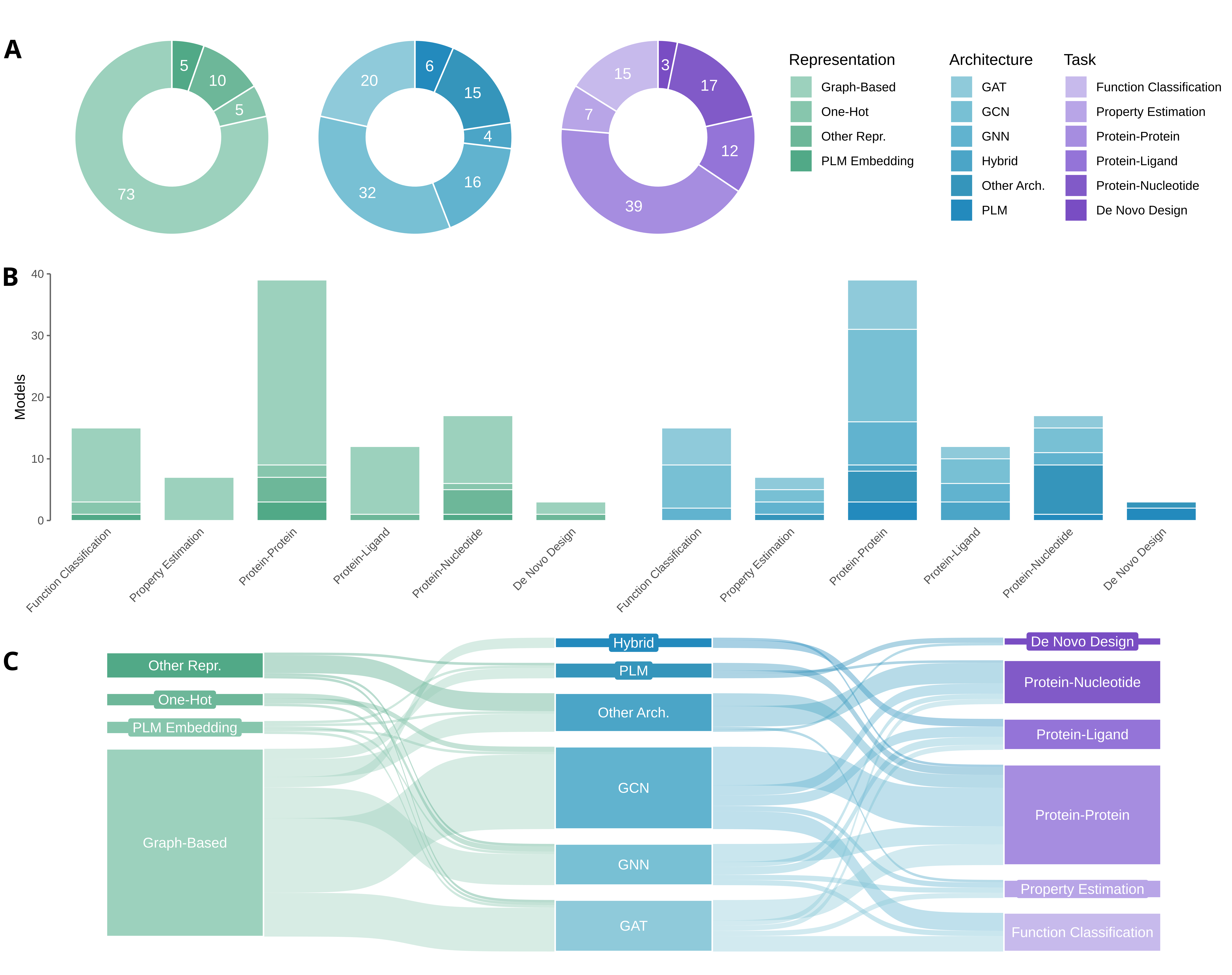}
    \caption{\textbf{Landscape of geometric deep learning (GDL) models applied to protein-related tasks}. This figure summarizes more than 90 GDL-based models extracted from the literature and analyzed in this perspective. \textbf{A.} The donut charts represent the overall distribution of models categorized by input representation strategies (left-green), model architectures (center-blue), and targeted biological tasks (right-purple), where segment sizes reflect absolute counts. Graph-based representations (73\%) dominate over PLM embeddings, one-hot, and handcrafted features. Among architectures, GCNs and GATs are the most frequently employed, while protein-protein interaction prediction emerges as the most common task. \textbf{B.} Bar plots illustrate the distribution of architectures and representations employed across six most general applications detected in this perspective. This highlights that protein-protein tasks are heavily studied and leverage a wider range of architectures and input types, whereas \textit{de novo} design remains underexplored. \textbf{C.} A Sankey diagram showing the dominant modeling pipelines, connecting data representations (left), architectural families (middle), and target tasks (right). The flow widths are proportional to the number of models using each pathway. This visualization reveals that most models follow a path from graph-based representations to GCN/GAT architectures, culminating in protein-protein prediction or function classification tasks, whereas combinations involving PLM-based embeddings and hybrid architectures are less common but indicate growing interest.}
    \label{fig:figure_03}
\end{figure}

To better understand the methodological and conceptual landscape of GDL in protein science, Figure~\ref{fig:figure_03} presents a systematic analysis of more than 90 GDL-based models reported in the literature. The figure characterizes the distribution of these models according to input representation, neural architecture, and the specific biological task they aim to solve. The majority of models employ graph-based representations of molecular structure—often derived from protein backbones or residue contact maps—underscoring the prominence of structural encoding in GDL frameworks. Among architectural choices, GCNs and GATs dominate, likely due to their flexibility, scalability, and established success in molecular tasks.

In the following section, we provide a task-oriented overview of GDL applications in protein modeling, structured according to the major categories illustrated in Figure~\ref{fig:figure_03}. For each task, we review representative models, analyze methodological choices, and identify emerging trends and limitations that define the current state of the field.

\subsection{Prediction of Functional Activity}

Functional classification models employ GDL to extract meaningful biological signals from protein structure and sequence, enabling the prediction of a wide range of biochemical properties. For instance, sAMPpred-GAT constructs residue-level graphs from predicted peptide structures and utilizes graph attention layers to model inter-residue relationships that drive antimicrobial activity \citep{yan2023samppred}. Similarly, AMPs-Net processes atom-level graphs enriched with physicochemical attributes using GCNs to predict peptide functionality \citep{ruiz2022rational}. LABAMPsGCN adopts a distinct approach by modeling heterogeneous graphs that represent peptides and their subsequences, capturing hierarchical dependencies often missed by traditional methods \citep{sun2022labampsgcn}. In parallel, TP-LMMSG and ACP-GCN combine multiple pre-trained embeddings—derived from language models and structural features—to enrich the representational power of the model, though this strategy often increases computational complexity \citep{chen2024tp, rao2020acp}. These approaches underscore the growing role of multi-scale and multimodal representations in modern GDL frameworks for peptide function prediction \citep{wei2021atse}.

More recent models have pushed these boundaries even further by incorporating contrastive learning, pre-training, and structure-aware representations. For example, PepGB integrates fine-tuned peptide embeddings with a perturbation-aware module that enhances model sensitivity to local structural changes and is especially effective under class imbalance scenarios \citep{lei2024pepgb}. Similarly, the Multi-Peptide framework leverages PeptideBERT representations in conjunction with GNN encoders to fuse sequence-derived and structure-based features, improving predictive generalization across diverse peptide tasks \citep{guntuboina2023peptidebert}. GraphCPP focuses specifically on cell-penetrating peptides, modeling residue-level interactions using GNNs trained to identify membrane-associated motifs, and demonstrates how local graph context can improve prediction of peptide uptake \citep{imre2025graphcpp}.

In protein function prediction, several structure-based GDL models have shown that integrating AlphaFold2-derived protein structures with advanced graph architectures significantly boosts performance. DeepFRI exemplifies this by predicting Gene Ontology (GO) terms through a combination of structure-based GCNs and transformer-derived sequence embeddings, outperforming sequence-only models and scaling robustly across protein families \citep{gligorijevic2021structure}. Similarly, GAT-GO utilizes residue-level attention mechanisms to weight structural relevance in function assignment. Struct2GO and PANDA2 further refine this paradigm by combining predicted structures with amino acid-level and global features, allowing better generalization across species and biological contexts \citep{10.1093/bioinformatics/btad637, 10.1093/nargab/lqac004}. DeepGraphGO and DeepGATGO take a more integrative route by combining protein sequence, structure, and protein-protein interaction networks into a unified graph framework, while also modeling the GO label hierarchy through graph-based encodings \citep{10.1093/bioinformatics/btab270, li2023deepgatgo}. These models not only improve functional accuracy but also support cross-species prediction, a major challenge in computational biology.

Moreover, GDL has proven effective in modeling toxicity and multifunctionality. In toxicity prediction, tAMPer and ATSE merge sequence-based features with graph-derived structural encodings, enhanced by bidirectional recurrent networks and attention mechanisms that focus the model on toxicophoric substructures \citep{ebrahimikondori2024structure, wei2021atse}. These attention-based modules improve interpretability and allow the model to prioritize biologically relevant motifs. Similarly, multifunctional peptide predictors such as iMFP-LG and MFP-MFL extend GDL capabilities to multi-label classification tasks, using graph-based modeling of label correlations alongside ensemble learning to deliver robust predictions across functional categories, including antimicrobial, anticancer, and antiviral peptides \citep{luo2024imfp, ge2025mfp}.

These developments reflect a broader trend in the field: the convergence of sequence, structure, and label ontology into unified GDL models. Whether through hierarchical label graphs, attention mechanisms, or multimodal embeddings, the next generation of predictive models increasingly relies on architectures that capture biological complexity in graph form. These strategies, while computationally intensive, offer unprecedented precision in characterizing peptide and protein behavior, and continue to reshape the landscape of bioinformatics and therapeutic peptide discovery.

\subsection{Prediction of Physicochemical Properties}

Geometric deep learning models targeting physicochemical property prediction commonly focus on estimating the impact of point mutations within protein sequences, including changes in stability and solubility. Models such as GLGNN-UCL employ siamese architectures to compare wild-type and mutant protein structures, directing attention to the mutated residues via localized attention mechanisms \citep{Gong2023GLGNN-UCL}. This architecture enables fine-grained prediction of changes in thermodynamic stability ($\Delta \Delta G$), facilitating the identification of stabilizing and destabilizing mutations \citep{Wang2023Pros-GNN}. ThermoAGT-GA and ProSTAGE extend this framework by incorporating enriched node features derived from pre-trained protein language models and by embedding edge representations that capture residue–residue orientations and spatial proximities \citep{madani2024, Li2024ProSTAGE}. These enhancements allow more expressive modeling of structural perturbations induced by mutations.

However, a central limitation of these approaches lies in their dependence on high-quality structural data for both the wild-type and mutant variants. In many cases, such structural information is unavailable or uncertain, particularly for proteins lacking homologous templates or for mutations that induce significant conformational shifts \citep{varadi2025challenges}. This raises important questions about the generalizability and robustness of current GDL frameworks in low-confidence structural contexts. Despite advances in structure prediction methods, the accuracy of mutant models remains variable and may constrain the applicability of these techniques in high-throughput or clinical settings.

In the context of solubility prediction, models such as DeepMutSol and GATSol integrate graph convolutional networks and attention mechanisms to combine sequence-derived features with structural graph representations \citep{Wang2024DeepMutSol, Li2024GATSol}. These multi-modal approaches show improved performance in capturing solubility changes induced by mutations, yet they also inherit susceptibility to inaccuracies in the predicted structures, which can introduce noise and confound downstream interpretations \citep{song2023accurately}. Addressing this issue may require the development of uncertainty-aware models or hybrid architectures that adaptively weight sequence and structure contributions based on input reliability.

More recent models, such as GPSFun, exemplify the scalability and versatility of GDL-based methods by leveraging deep GNN stacks to simultaneously predict multiple physicochemical properties from structural inputs \citep{Yuan2024GPSFun}. This direction highlights the potential of unified frameworks capable of modeling complex, interdependent biophysical traits. As the field advances, balancing architectural complexity with data availability and predictive robustness remains a key challenge—especially when translating these models to real-world protein design scenarios where experimental validation is limited or delayed.

\subsection{Predicting-Protein Intermolecular Interactions}

Understanding and predicting intermolecular interactions is fundamental to protein function and serves as a foundational element of applications in drug discovery, synthetic biology, and molecular diagnostics. GDL offers a powerful framework for modeling such interactions by capturing the spatial, topological, and physicochemical features that govern molecular recognition and binding \citep{MaSIF}. 

In this section, we highlight representative GDL strategies developed to model interactions between proteins and other biomolecules, including ligands and nucleic acids. These approaches demonstrate how GDL architectures enable the accurate prediction of binding interfaces, interaction affinities, and structural complementarity across diverse molecular systems. 

\subsubsection{Protein–Protein Interactions}

Protein–protein interaction (PPI) prediction has seen significant progress through the application of GDL, extending from the identification of binding sites to the modeling of changes in binding affinity. Early models such as GraphPPepIS and AGAT-PPIS utilize graph-based representations enriched with residue-level or atomic features, capturing the topological and physicochemical context of interfaces \citep{li2023simultaneous, zhou2023agat}. AGAT-PPIS further incorporates evolutionary profiles to enhance biological realism and context-awareness, improving predictions particularly in homologous or functionally conserved interaction settings \citep{li2023simultaneous, zhou2023agat}. 

Building upon these foundations, models like EGRET and DeepProSite introduce deeper architectures through transformer-based modules and graph attention mechanisms that explicitly model long-range dependencies and interface-specific patterns \citep{mahbub2022egret, fang2023deepprosite}. These enhancements illustrate the strength of GDL in resolving the spatially constrained and chemically complex nature of protein interfaces. Nonetheless, scalability remains a pressing concern, as the computational cost and memory footprint grow substantially with the size of the interacting complexes. This poses limitations for modeling large assemblies such as ribosomes or viral capsids, where traditional GDL models may not scale efficiently.

Affinity prediction frameworks have followed a similar trajectory, with models like PASNVGA and DDMut-PPI adopting siamese GNN designs to explicitly compare wild-type and mutant complexes \citep{luo2023predicting, zhou2024ddmut}. By aligning embeddings of interacting partners before and after mutation, these models provide fine-grained estimations of changes in binding strength, enabling prioritization of mutations that enhance or weaken affinity \citep{luo2023predicting, zhou2024ddmut}. 

Complementary strategies such as MM-StackEns and MEG-PPIS implement ensemble learning and multiscale geometric encodings to better generalize across diverse protein pairs, mitigating overfitting and capturing hierarchical structural cues \citep{albu2023mm, ding2024meg}. These approaches are particularly valuable in settings where training data are limited or derived from structurally heterogeneous sources.

Furthermore, GDL models targeting peptide–protein interactions, such as PepNN-Struct and PeSTo, highlight the growing importance of flexibility-aware architectures \citep{abdin2022pepnn, PeSTo}. These models integrate reciprocal attention mechanisms and geometric transformers to accommodate backbone variability and dynamic binding modes, traits that are especially relevant in signaling peptides, disordered motifs, or peptide-based therapeutics \citep{abdin2022pepnn, PeSTo}. By explicitly modeling structural plasticity, these architectures extend the applicability of PPI predictors to less structured and more flexible systems, which are often underrepresented in conventional datasets.

Recent advances in protein–protein interaction modeling using geometric deep learning indicate a transition toward architectures capable of capturing detailed structural features while also incorporating evolutionary, dynamic, and contextual information. Despite these advances, several limitations remain, including issues related to computational scalability, the interpretability of model predictions, and the integration of structural data that may be incomplete or uncertain. Addressing these challenges may involve the adoption of hybrid strategies that integrate geometric deep learning with coarse-grained approximations or uncertainty-aware frameworks, with the goal of maintaining predictive performance while extending applicability to complex or experimentally under-characterized systems.

\subsubsection{Protein–Ligand Interactions}

Geometric deep learning approaches for modeling protein–ligand interactions have evolved substantially, transitioning from early voxel-based convolutional architectures to more refined graph-of-graphs representations that better capture the complexity of biochemical binding. Foundational models such as KDEEP and GraphDTA marked a changing toward spatially informed prediction of binding affinities by integrating structural and topological features into supervised learning frameworks \citep{jimenez2018k, nguyen2021graphdta}. Building on this, GEFA introduced attention-guided message passing to explicitly model contextual interactions between ligand atoms and protein residues, improving sensitivity to functional site environments \citep{nguyen2021gefa}.

More recent architectures, such as PIGNet2, incorporate physically grounded interaction terms—including van der Waals forces and hydrogen bonding potentials—into the graph learning process, enabling direct modeling of energetic contributions and structural rearrangements across the ligand–protein interface \citep{moon2024pignet2}. These enhancements have led to improved prediction of thermodynamic quantities such as binding free energy, underscoring the value of integrating inductive biases from molecular physics into GDL pipelines.

Multimodal architectures have also gained prominence. EmbedDTI and CPI-GGS exemplify efforts to bridge sequence-based and graph-based representations, typically employing convolutional neural networks for encoding protein sequences and graph neural networks for small molecule representations \citep{jin2021embeddti, hou2025cpi}. Attention mechanisms serve as the bridge between these modalities, enabling the alignment of heterogeneous embeddings and facilitating end-to-end learning across domains \citep{jin2021embeddti}. While these models offer greater versatility, they remain constrained by their reliance on sequence-level protein features, which may lack the spatial resolution required to model highly specific binding events or conformational rearrangements.

The introduction of frameworks like EGGNet signals a further refinement in representational capacity. By adopting equivariant GNNs and hierarchical graph-of-graphs encodings, EGGNet accommodates flexible binding scenarios involving non-canonical residues, macrocycles, or allosteric modulators \citep{wang2024eggnet}. Its architecture is well-suited for generalization across structurally diverse chemical scaffolds and complex protein folds. However, a persistent challenge in this domain is the limited availability and variable quality of structural data—both for proteins and their bound ligands. Predicted structures, particularly those generated for low-resolution or highly dynamic targets, may introduce artifacts or noise that propagate through the model and degrade prediction accuracy \citep{libouban2023impact}.

Moreover, many current models operate under the assumption of well-annotated ligand binding poses and complete interaction graphs, which do not always reflect the sparsity or ambiguity of experimental datasets. The lack of standardized benchmarking for real-world docking scenarios further complicates performance assessment. Future advances may benefit from integrating uncertainty quantification, active learning, or flexible alignment modules that adaptively adjust to input quality. As the field continues to expand, developing robust architectures that maintain predictive power in structurally noisy or low-confidence environments remains an essential frontier for GDL-guided drug discovery \citep{morehead2025deep}.

\subsubsection{Protein–Nucleotide Interactions}

GDL approaches applied to protein–nucleotide interactions have steadily advanced from simple spatial contact prediction to more sophisticated frameworks for binding site classification. Early models such as GraphSite and GLMSite construct distance-based residue graphs from predicted structures—most often derived from AlphaFold or ESMFold—and use spatially encoded embeddings to pinpoint DNA- and RNA-binding regions with notable accuracy \citep{yuan2022alphafold2, song2023accurately}. These models have demonstrated strong performance on benchmark datasets, capturing conserved interface patterns across protein families. However, they often rely on rigid structural assumptions and fail to fully account for the conformational flexibility of nucleic acids.

More recent developments have sought to address the modality gap between structured ligands and protein surfaces. ZeRPI, for instance, introduces a contrastive learning framework to align RNA graphs with protein embeddings, enabling zero-shot prediction of interaction sites even in the absence of labeled training pairs \citep{gao2025zerpi}. This cross-modality transfer capacity reflects a growing emphasis on model generalizability in data-scarce regimes, particularly for underrepresented RNA types and noncanonical secondary structures.

Further innovations such as EquiPNAS incorporate equivariant neural networks informed by pre-trained protein language models, capturing residue-wise orientation and symmetries relevant to nucleic acid binding \citep{roche2024equipnas}. GeoBind shifts the modeling paradigm from atomic graphs to surface point clouds, applying geodesic convolutions that preserve local curvature and topology of protein binding pockets \citep{li2023geobind}. Similarly, EGPDI combines multiple graph views—including structural and sequence-derived features—to enhance predictive robustness across diverse protein–DNA complexes \citep{zheng2024egpdi}. These approaches collectively emphasize the versatility of GDL frameworks in capturing nuanced features of protein–nucleotide interactions.

Despite this progress, relevant challenges persist. One of the most pressing limitations is the difficulty in modeling flexible nucleotide geometries and rare structural motifs—features that are essential for capturing biologically relevant interfaces in transcription factors, ribozymes, or RNA-guided complexes \citep{tarafder2024landscape}. Many current models rely heavily on high-confidence structural data, which are often unavailable or unreliable for proteins with low structural coverage or intrinsically disordered regions. Moreover, benchmark datasets used for training and evaluation frequently omit noncanonical nucleotides or dynamic binding modes, thereby limiting the scope of model generalization \citep{sabei2024dynamics}.

Efforts to extend these models to more realistic settings will likely require architectures that can account for conformational ensembles and uncertainty in both protein and nucleotide states. The incorporation of attention-based modules, equivariant transformations, and ensemble-based or probabilistic representations may help mitigate noise and enhance interpretability. At the same time, active learning strategies and adaptive alignment mechanisms may offer improved resilience in low-resource settings, allowing models to remain performant even when structural inputs are ambiguous or incomplete \citep{lu2025aligning, vinchurkar2025uncertainty}

While current GDL methods have demonstrated impressive accuracy under idealized conditions, achieving robust performance in the face of structural heterogeneity and nucleic acid flexibility remains an open frontier. Bridging this gap will be essential for enabling GDL-based predictors to function effectively in systems biology and RNA-targeted drug discovery.

\subsection{\textit{De Novo} Protein Design}

Generative GDL models aim to create novel proteins with predefined structural or functional attributes. Frameworks like RFdiffusion and ProteinMPNN employ message-passing and diffusion strategies to iteratively construct sequences compatible with given backbones \citep{Rfdiffusion, RFDiffusionAA, ProteinMPNN}. These methods offer precision and flexibility, though they often require detailed structural templates and face challenges in handling variable-length sequences \citep{baek2021accurate}.

CARBonAra and Genie leverage point cloud and reference frame models to preserve rotational and angular information, enhancing generation fidelity \citep{CARBonAra}. FrameDiff and PROTSEED further utilize Invariant Point Attention to update node features iteratively, improving structural consistency across generations. Such methods demonstrate excellent structural alignment but are computationally intensive \citep{FrameDiff, PROTSEED}.

Graph-based approaches like ADesign and GVP-GNN use deep GNN encoders to represent macromolecular context, predicting sequences that fit 3D scaffolds \citep{gao2022alphadesigngraphproteindesign, GVP-GNN}. GeoSeqBuilder and ProteinSolver implement convolutional graph layers to model folding constraints \citep{GeoSeqBuilder, ProteinSolver}. Meanwhile, antibody-focused models like RefineGNN and AbODE exploit hierarchical message passing and partial differential equations to navigate the complex topologies of antigen-binding sites \citep{RefineGNN, AbODE}.

These diverse methodologies illustrate GDL’s broad applicability in protein design. However, they also underscore ongoing limitations such as high data requirements, interpretability concerns, and computational scalability. Nonetheless, GDL continues to evolve, offering powerful, structure-aware strategies that push the boundaries of rational and autonomous protein engineering.

\section{Challenges and Perspectives on the Integration of Geometric Deep Learning for solving Protein Engineering Problems}

The integration of GDL in protein engineering has opened new oportunities for understanding, predicting, and designing complex biomolecular systems. Yet, as with any transformative technology, its incorporation comes with both opportunities and limitations. 

This section discusses the major strengths and weaknesses of GDL, addresses the technical and conceptual challenges associated with its use in protein science, and outlines methodological strategies for its effective deployment. Emphasis is also placed on the central issue of explainability, a key requirement for adopting artificial intelligence models in biologically and clinically relevant contexts.

\subsection{Strengths and Limitations of GDL in Protein Engineering}

Geometric deep learning offers distinct advantages for modeling proteins, whose functions are intricately linked to their three-dimensional spatial organization \citep{meng2025protein}. By leveraging non-Euclidean representations, GDL facilitates the integration of structural, chemical, and functional features within a unified framework \citep{GVP-GNN}. This capability has translated into superior performance across a range of predictive tasks, including binding site identification, stability estimation, and even \textit{de novo} protein sequence generation \citep{GeoSeqBuilder, CARBonAra, MaSIF}.

A relevant strength of GDL lies in its flexibility to accommodate different molecular representations. Models can operate at multiple resolutions—residue-level, atomic-level, or using coarse-grained abstractions—depending on the task and available data \citep{RFDiffusionAA}. Architectures such as GVP-GNN and RFdiffusion exemplify how geometric encoders can faithfully capture molecular symmetries and geometric constraints, outperforming traditional sequence-based models in tasks that require structural sensitivity \citep{GVP-GNN, Rfdiffusion}. This structural awareness enables improved generalization to novel folds, rare variants, or out-of-distribution mutations, where purely sequence-based models often fail \citep{Balaji2025}.

Despite these strengths, several limitations hinder the broader applicability of GDL in protein science. First, these models are inherently data-intensive and often require high-resolution, experimentally derived structures, which remain scarce for unstable, disordered, or rare protein families \citep{wu2023integration}. Moreover, the computational demands of training and inference—particularly for deep architectures or large molecular graphs—pose a challenge for high-throughput applications \citep{Sverrisson2020.12.28.424589}.

From a methodological standpoint, GNN-based models exhibit technical vulnerabilities that can compromise biological fidelity. Repeated message passing can lead to oversmoothing, wherein node embeddings converge to indistinct representations, eroding the model’s capacity to resolve local functional sites \citep{song2023ordered}. At the same time, deep architectures often suffer from message-passing bottlenecks, which restrict the flow of global context in heterogeneous protein graphs. Over-squashing—where complex information is compressed into low-dimensional representations—further impairs both accuracy and interpretability \citep{akansha2025over}. To mitigate these effects, recent architectural innovations have incorporated residual connections, hierarchical pooling layers, and attention-based mechanisms to preserve resolution across spatial scales and maintain task-relevant specificity \citep{wu2025hierarchical}.

Despite promising performance \textit{in silico}, integrating GDL models into experimental and clinical workflows remains a challenge. Many models lack the interpretability, uncertainty quantification, or robustness required to support high-stakes biological decisions, such as drug target validation or mutational risk assessment \citep{lavecchia2024advancing}. Bridging this gap will require not only algorithmic refinement but also improved model transparency and better alignment with domain-specific constraints \citep{medina2024interpretable}.

\subsection{Technical and Conceptual Challenges}

Among the foremost challenges in GDL-assisted protein engineering is the issue of data acquisition and standardization. While models such as AlphaFold2 and ESMFold have made it possible to predict high-quality protein structures, experimental validation remains essential \citep{Jumper2021IPA, lin2023evolutionary}. Moreover, discrepancies between predicted and experimentally observed conformations can propagate errors through GDL models, especially when fine-grained functional predictions are required \citep{lin2023evolutionary}.

Graph construction is another bottleneck. Defining appropriate node and edge features—such as using C$\alpha$ distances, biochemical attributes, or predicted contacts—requires careful tuning \citep{muller2024a}. Choices in graph granularity (e.g., residue-level vs. atom-level) significantly affect performance and computational burden \citep{Coarse-Grained_Review}. The use of automated pipelines, such as those provided by Graphein, can streamline this step but may introduce variability \citep{jamasb2022graphein}.

Model training also presents specific obstacles. Limited annotated datasets for certain tasks, such as allosteric modulation or protein–RNA interactions, hinder supervised learning \citep{Khakzad2023, zhu2025identifying}. Strategies such as data augmentation, transfer learning from language models, or unsupervised pretraining offer promising alternatives \citep{Sun2024_DataAug, wu2023integration}. Training protocols must also address class imbalance, overfitting, and scalability \citep{Jiang2025_ReviewImbalance}. Frameworks such as semi-supervised learning and multi-task learning may offer improved generalization in such scenarios \citep{dai2025surveydeeplearningmethods}.

The design of GDL models should be closely aligned with their intended application. In rational design scenarios, where one starts with a known structure and seeks to optimize a specific region or function, GDL can assist in identifying hotspots or favorable mutations \citep{Wang2024DeepMutSol, Wang2023Pros-GNN}. In ML-guided directed evolution\citep{yang2019machine}, GDL models can prioritize variants for experimental testing by predicting fitness landscapes or mutational effects \citep{cheng2024zero}. In \textit{de novo} design, GDL will contribute by generating sequence-structure pairs that satisfy geometric or functional constraints, accelerating the path to novel folds or functions \citep{Rfdiffusion, Chroma, CARBonAra}.

\subsection{Explainability in Geometric Deep Learning for Protein Modeling}

Despite the rapid advances in GDL, particularly through graph neural networks, the interpretability of these models remains a fundamental and unresolved challenge \citep{yu2021towards}. Their black-box nature obscures the rationale behind predictions—an issue of particular concern in high-stakes biomedical applications such as protein engineering, drug discovery, and clinical decision-making, where transparency, trust, and regulatory accountability are paramount \citep{medina2024interpretable}. While convolutional neural networks have benefited from a mature ecosystem of explainability tools, the field of graph-based explainable AI is comparatively emergent, especially in the context of molecular design \citep{nandan2025graphxai}.

Recent efforts have begun to address this interpretability gap. Post hoc methods for GNNs can be broadly categorized into gradient-based techniques, perturbation-based methods, surrogate modeling, subgraph generation, and counterfactual explanations \citep{he2024explaining, nandan2025graphxai}. Gradient-based approaches—such as GraphGrad-CAM \citep{yu2021towards}, GNN-LRP \citep{schnake2021higher}, DeepLIFT \citep{shrikumar2017learning}, and integrated gradients—trace the flow of gradients through a trained model to attribute prediction relevance to specific nodes or features. Perturbation-based strategies, including PGExplainer \citep{luo2020parameterized} and SubgraphX \citep{yuan2021explainability}, evaluate how localized changes in the graph topology affect model outputs, assigning probabilistic importance to edges or substructures. Frameworks like GraphLIME \citep{huang2022graphlime} extend local surrogate modeling to graph domains, while GraphSVX and PGM-Explainer build interpretable approximations of the decision boundary using supervector decomposition or probabilistic graphical models, respectively \citep{duval2021graphsvx, vu2020pgm}.

Subgraph-based explanations construct compact graphical motifs that are sufficient to drive a specific prediction. Methods such as XGNN employ reinforcement learning to generate class-specific subgraphs \citep{yuan2020xgnn}, while IFEXPLAINER utilizes information flow metrics to isolate causal structures \citep{yu2021towards}. Counterfactual techniques—such as CF-GNNExplainer \citep{lucic2022cf}—identify minimal structural changes that would alter a prediction, offering insight into model robustness and decision boundaries.

In parallel, interpretability techniques are increasingly embedded into the architecture of GDL models. Attention-based mechanisms, such as those used in Graph Attention Networks \citep{Velikovi2017} and ASAP \citep{ranjan2020asap}, assign dynamic weights to neighboring nodes during message passing, facilitating intuitive visualization of influential residues or interactions. Other intrinsic methods incorporate information bottlenecks \citep{musat2022information}, causality-aware frameworks \citep{behnam2024graph}, or self-explaining modules designed to prioritize interpretability alongside accuracy \citep{zhang2025interpretable}. However, while attention weights are accessible and visually appealing, their consistency and biological relevance can be unreliable, varying across training runs and random seeds \citep{hao2021self}.

In protein modeling, interpretability tools have been applied to identify catalytically active residues, conserved motifs, or flexible regions relevant to binding or stability. For example, attention maps from GATs have highlighted surface residues aligned with experimental binding sites, while perturbation-based approaches such as GNNExplainer have recovered subgraphs corresponding to secondary structure elements implicated in functional specificity \citep{gruver2023protein, li2022mfnet}. These visualization strategies are often overlaid on molecular surfaces, aiding intuitive interpretation and informing decisions related to residue truncation, mutagenesis prioritization, or structural refinement \citep{medina2024interpretable}.

Yet, limitations persist. Most current techniques provide only local, instance-specific explanations that do not generalize across protein families or structural classes. Attribution methods often fail to capture the true causal determinants of molecular function and can produce unstable outputs that are sensitive to hyperparameters and model initialization. In large, densely connected graphs, such explanations may become overly complex and biologically opaque, undermining their practical utility \citep{kakkad2023survey}.

To overcome these challenges, recent studies emphasize the integration of structural priors, evolutionary constraints, and domain-specific heuristics to constrain explanation generation. Combining attention-based layers with post hoc attribution techniques can yield more consistent and biologically plausible interpretations \citep{lopardo2024attention, panagiotaki2023semantic}. Moreover, generative and counterfactual frameworks enable the exploration of minimal mutations or structural rearrangements that transform function, offering mechanistic insights that extend beyond traditional classification or regression outputs \citep{saranti20243d}.

Explainable GDL methods are particularly powerful when embedded within iterative design pipelines. In this setting, interpretability transcends its conventional post hoc role to become an active driver of hypothesis generation, experimental prioritization, and model refinement. For instance, identifying subgraph motifs associated with thermostability can inform targeted residue substitutions, while highlighting latent determinants of activity can guide sequence diversification in directed evolution. When systematically leveraged, these insights contribute not only to more transparent modeling but also to more efficient and rational protein engineering.

While explainability in GNNs remains an open and evolving frontier, its potential to transform protein science is profound. Continued innovation in this space will be crucial for developing interpretable, trustworthy, and biologically grounded deep learning tools that can accelerate discovery and design in the molecular sciences.

\section{Integrating Geometric Deep Learning into Protein Design Pipelines}

\begin{figure}[!htpb]
    \centering
    \includegraphics[width=\linewidth]{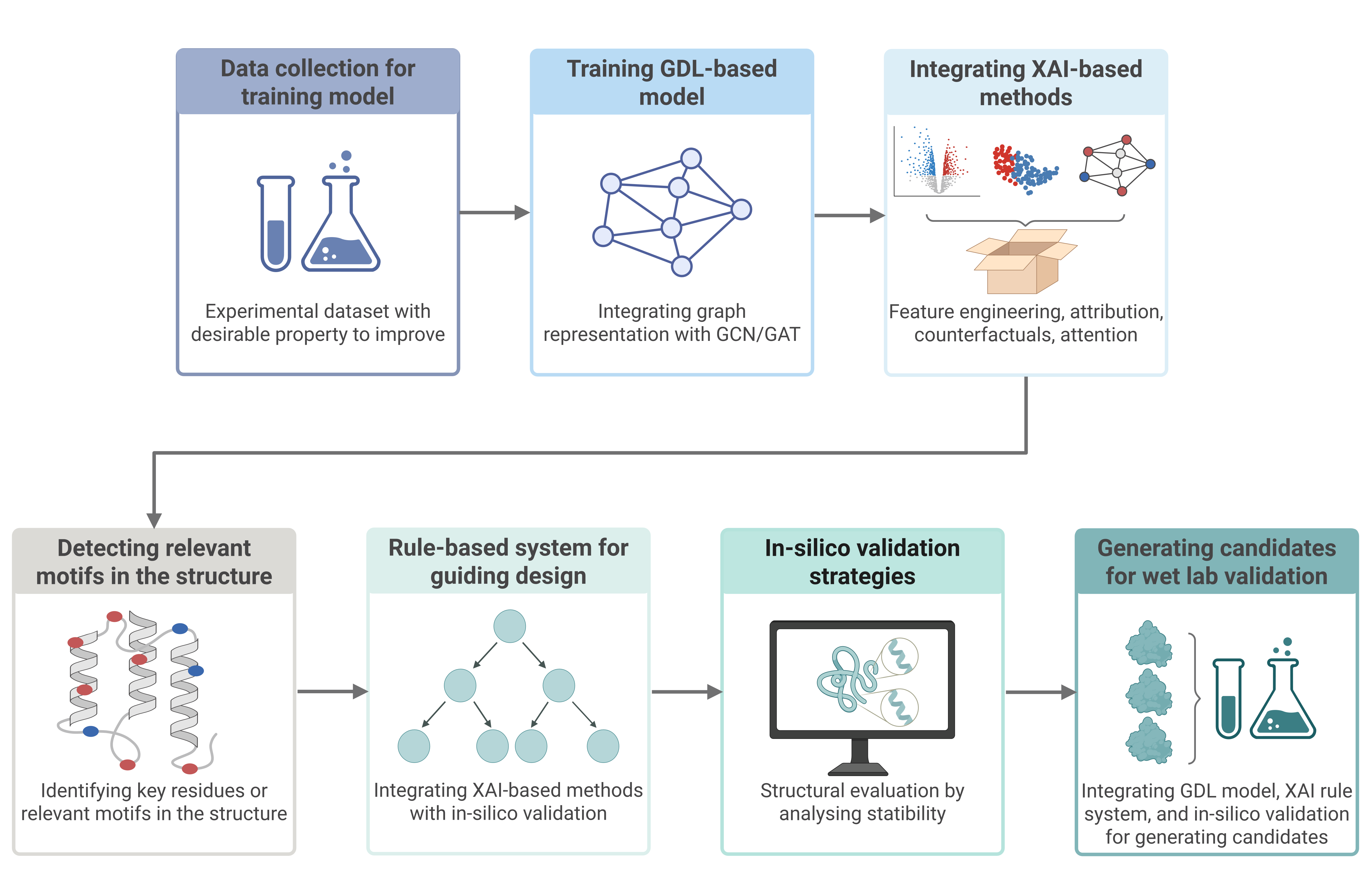}
    \caption{\textbf{Overview of an integrative pipeline for protein design combining geometric deep learning and explainable artificial intelligence}. The workflow begins with the collection of experimental datasets from mutational libraries or high-throughput assays, which are used to train GDL models based on graph representations. Explainability is incorporated through feature attribution, counterfactual analysis, and attention mechanisms to identify key structural motifs and mutational hotspots. These insights are synthesized into a rule-based system that guides the design process and prioritizes candidates. Predicted variants are then evaluated using in silico validation strategies—including stability estimation and structural assessment—to generate biophysically plausible candidates for wet-lab testing.}
    \label{fig:figure_04}
\end{figure}

The integration of GDL into protein design workflows represents a critical milestone toward enabling interpretable, structure-aware, and biologically grounded computational pipelines \citep{medina2024interpretable}. While previous sections have explored predictive and generative strategies across diverse protein tasks, here we propose a unified framework that consolidates these advances into a cohesive and scalable design pipeline (Figure \ref{fig:figure_04}). This section outlines the methodological integration of GDL models, XAI, and bioinformatic validation tools, focusing on the application to protein engineering and mutational variant selection.

The pipeline begins with the systematic collection of experimental datasets, typically derived from protein mutational libraries or high-throughput assays that quantify catalytic activity, binding affinity, or general fitness-related properties. These datasets form the inputs for supervised learning tasks and guide subsequent predictive efforts.

Following data collection, structural information for each protein variant is either obtained from high-resolution experimental sources or predicted using state-of-the-art models. These predicted structures serve as the input for training a GDL-based predictive model, as described in Section \ref{sec:sec-implementing}. 

Explainability is embedded into the pipeline through a multi-step strategy. First, feature engineering is used not only for input representation but also for retrospective analysis of model predictions. Techniques such as permutation importance and gradient-based allow the identification of features that most influence the predicted fitness \citep{montavon2018methods}. In parallel, counterfactual reasoning is implemented by generating graph variants through systematic perturbations at the residue or edge level \citep{nandan2025graphxai}. These modified graphs are used to observe changes in the predictive output, enabling the identification of critical mutations with positive or negative effects on protein fitness.

Attention-based mechanisms further enhance interpretability by assigning dynamic weights to residues and interactions during message passing \citep{hao2021self}. These weights help highlight structurally or functionally relevant regions, which can be mapped onto the three-dimensional structure for visual and mechanistic insights. Complementing this, subgraph discovery techniques identify minimal motifs sufficient for prediction—often corresponding to active sites, stability cores, or conserved domains—thus capturing the underlying structural logic of the model's decision \citep{yuan2021explainability, medina2024interpretable}.

To synthesize these diverse insights, a decision rule-based module aggregates outputs from attribution, attention, and counterfactual layers to assign probabilistic relevance scores to regions of interest. This aggregation enables prioritization of mutational hotspots based on estimated success likelihood. Additionally, uncertainty-aware modeling—implemented through variational or Bayesian GDL frameworks—provides confidence estimates that are critical for prioritizing experimental validation \citep{wang2022bayestab}.

The predictive candidates are then subjected to post hoc filtering using physics-based simulations, such as energy minimization or structural relaxation, allowing biophysically plausible variants to be retained. This process is complemented by structural validation tools, which assess secondary structure preservation, torsional constraints, and energetic favorability.

\subsection{Methodological Innovations and Key Challenges}

Several novel elements distinguish this pipeline. First, it closely integrates GDL with XAI methods, enabling mechanistic insights into sequence–function relationships. Second, the use of counterfactuals and subgraph attribution promotes model transparency, essential for experimental trust and decision-making \citep{nandan2025graphxai, medina2024interpretable}. Third, probabilistic modeling provides uncertainty-aware predictions, critical for real-world deployment \citep{wang2022bayestab}.

Despite these advances, major challenges remain. Low-data regimes (\textit{Low-N}) limit the training of deep models, especially for rare proteins or newly engineered enzymes \citep{biswas2021low}. Transfer learning from pretrained protein models and contrastive self-supervision offer potential solutions, but their integration into GDL remains underexplored \citep{BARBEROAPARICIO2024102035}. Additionally, experimental validation introduces latency and cost, necessitating accurate prioritization mechanisms—an area where uncertainty quantification and XAI convergence will be particularly valuable \citep{vinchurkar2025uncertainty}.

Moreover, the capacity for generalization across protein families and structural folds remains an open frontier. Future efforts should include domain adaptation techniques, cross-clade pretraining, and homology-aware validation protocols. Extending GDL workflows beyond sequence proximity and toward functional generalization will be critical for their broader applicability \citep{rao2019evaluating}.

\section{Future Directions and Transformative Potential of GDL in Protein Engineering}

GDL is rapidly emerging as a foundational paradigm in next-generation protein engineering. By bridging structural biology with advanced artificial intelligence, GDL enables a direct engagement with the molecular geometry that underlies biological function. Its capacity to capture spatial and topological features with high fidelity has proven invaluable across a broad spectrum of tasks, including predictive modeling, interaction analysis, and \textit{de novo} protein design.

This perspective has presented a comprehensive and critical overview of the current landscape of GDL applications in protein science. We have delineated its principal areas of impact—from the prediction of stability, solubility, and function, to the modeling of protein–ligand and protein–nucleotide interactions, and the emergence of generative frameworks for structural design. At the same time, we have highlighted methodological challenges that must be addressed for these models to achieve reliable, generalizable, and biologically grounded performance. Key among these are the imperatives of interpretability, reproducibility, and the principled incorporation of domain knowledge—features essential not only for scientific rigor but also for translational success.

GDL is playing an increasingly central role in protein engineering, particularly through its integration with emerging technologies such as pretrained protein language models, molecular dynamics simulations, high-throughput screening, and laboratory automation. Of particular promise are hybrid architectures that merge the structural resolution of GDL with the generative capacity of deep probabilistic models, enabling the design of novel proteins that satisfy both geometric and functional constraints. As curated datasets grow and interpretability techniques mature, GDL will serve as the computational backbone of closed-loop design cycles that unify in silico modeling with experimental synthesis and validation.

To realize this vision, future developments must prioritize transparency, robustness, and accessibility. Designing models that not only predict but also explain—and that can quantify uncertainty in their predictions—will be vital for their adoption in real-world protein engineering settings. Equally important is the promotion of interdisciplinary collaboration among computational scientists, structural biologists, chemists, and engineers, ensuring that methodological innovation is aligned with experimental feasibility and biological relevance.

GDL is more than a computational tool; it represents a conceptual shift in how we understand and manipulate the molecular machinery of life. By embedding structural principles at the heart of machine learning, GDL equips us with the means to predict, interpret, and design proteins with unprecedented precision. Its full potential lies not only in its algorithms, but in the collaborative systems it enables—systems capable of accelerating discovery, informing mechanism, and reshaping the future of synthetic biology.

\section*{Conflict of interest statement}

The authors declare no conflict of interests.

\section*{Author contributions statement}

JG-V, JR, NS-G, NM, DA-S, and DM-O: conceptualization. JG-V, JR, NS-G, NM and DM-O: investigation. JG-V, JR, NS-G, NM, MDD, CC, RU-P, JA, and DM-O: writing, review and editing. JA and DM-O: supervision and funding resources. JA and DM-O: project administration. All authors have read and agreed to the published version of the manuscript.

\section*{Acknowledgments}

DM-O and NS-G acknowledge ANID for the project “SUBVENCIÓN A INSTALACIÓN EN LA ACADEMIA CONVOCATORIA AÑO 2022”, Folio 85220004. DM-O, CC, RU-P, and JA gratefully acknowledge support from the Centre for Biotechnology and Bioengineering - CeBiB (PIA project FB0001 and AFB240001, ANID, Chile). DM-O acknowledges ANID for FONDECYT INICIACIÓN project 11250295. MDD acknowledges funding by the Deutsche Forschungsgemeinschaft (DFG, German Research Foundation) - within the Priority Program Molecular Machine Learning SPP2363 (Project Number 497207454). 

\bibliographystyle{apalike}

\end{document}